\documentclass[12pt]{article}
\usepackage{epsfig}
\begin{document}

\begin {center}
{\bf  {MOND - A REVIEW}}
\vskip 5mm
{\rm D.V. BUGG \footnote {email: david.bugg@stfc.ac.uk}} \\
{ Department of Physics, Queen Mary, University of London, London E1\,4NS,
UK} \\
\vskip -1mm
\end {center}

\begin{abstract}
\noindent
A critical appraisal is presented of developments in MOND since its introduction by Milgrom in 1983 to the present day.
\end{abstract}

\section {Introduction}
It is not possible to reference all papers on MOND. 
The discussion here follows the historical perspective except in a few places where later papers provide explanations of details in 
earlier ones. 

Rotation curves of galaxies disagree with Newton's laws.
The standard cosmological model $\Lambda CDM$ interprets this in terms of Cold Dark Matter.
There is a wide range of proposals what this Dark Matter may be. 
However, a long series of experiments has not located significant evidence for it. 
MOND (Modified Newtonian Dynamics) is a competing empirical scheme.
Famaey and McGaugh \cite {Famaey} have produced a review of the data up to 2011 with 
extensive references, which will be updated here.

By 1983, Tully and Fisher \cite {Tully} had established that rotation curves of galaxies over a
wide range of masses $M$ have asymptotic rotational velocities $V_\infty = (GMa_0)^{1/4} $;
$G$ is the gravitational constant and $a_0$ an empirical constant.
Masses are derived assuming $a_0$  is proportional to luminosity $L$. 
Also Faber and Jackson \cite {Faber} had studied the velocity dispersion $\sigma$ of random
motion of stars in high surface brightness elliptical galaxies and shown that  $\sigma^4 \simeq MGa_0$.
Milgrom based his scheme on these observations. 

He pointed out in his first paper \cite {MilgromA} that flatness of rotation curves remains constant down to radii well within galaxies.
He parametrised the full rotational acceleration $a$ in terms of the Newtonian acceleration $g_N$ as
\begin {equation}
a = g_N/\mu (\chi);
\end {equation}
$\mu$ is an empirical smooth function of $\chi = a/a_0$, where $a_0$ is a universal constant 
$\sim 1.2 \times 10^{-10}$ m s$^{-2}$ for all galaxies; $\mu \to 1$ for accelerations $\gg a_0$ and  for $a/a_0$, $\mu (x) \propto x$.
Three forms are in common use, and all go to  $ \sqrt {a_0g_N}$ as $g_N \to 0$.
From these relations, a star with rotational velocity $v$ in equilibrium with centrifugal force has
\begin {equation}
v^2/r = \sqrt {a_0 GM/r^2};
\end {equation}
the factor $r$ cancels and $v^4 = MGa_0$, the Tully-Fisher relation.

Milgrom's paper comments on the fact that the Milky Way has the shape of a thin disc, but it is unnecessary to account for motions
of stars in the $z$ direction, perpendicular to the disc.
These were studied by Oort \cite {Oort}.
The $z-$ excursions of stars are small compared with the orbital radius and deviations of velocities from circular motion are small.
One can determine $g_N$ and then via Poisson's equation find the total gravitational mass density in the central plane or
the surface mass density in the disc. 

Milgrom suggested that deviations from Newton's law may arise from variations of inertial mass.
He was led to this conclusion by the fact that within a star or nucleus, Newton's law applies, while on the
scale of a galaxy, the Tully-Fisher relation is required.
Later, in 1997, he introduduced the idea that equations of motion are invariant under the conformal
transformation $(t, \bf {r}) \to (\lambda t, \lambda \bf {r})$ in the limit of weak gravity; radii and accelerations
change under scaling, but velocities do not. 
The derivation is from Poisson's equation for a 2-D distribution like a disc galaxy, but is rather technical
\cite {Milgrom97}.

In 1984, Beckenstein and Milgrom considered a non-relativisitic potential $\phi$  for gravity differing from
Newtonian gravity using a Lagrangian formalism \cite {Beckenstein}.   
They derived a Poisson equation from this Lagrangian.
One result appeared from this work which would later prove to be very important.
Using the approximation that $\mu (x) \propto x$ for small $x$, they derived the result that
\begin {equation}
\phi \to \sqrt {GMa_0} \ln (r/r_0) + 0(r^{-1}),
\end {equation}
where $r_0$ is an arbitrary radius.
This leads to the asymptotically constant rotational velocity $V_\infty = (MGa_0)^{1/4}$.
They also considered the center-of-mass motion of two bodies, an issue which later became
important for QUMOND.

Further progress was slow while astrophysicists accumulated statistics on galaxies and
examined systematics.
Meanwhile studies were made with collaborators using models of galaxies.  
In 1996, Milgrom \cite {action} considered the virial equation for theories governed by an action, 
again laying the foundations for later developments.
The same year, Milgrom \cite {MilgromFJ} studied low surface density galaxies and accounted for the Faber-Jackson
relation $\sigma ^4 \propto MGa_0$ for the mean velocity dispersion $\sigma$ of self-gravitating systems supported by random motions.
This later proved important in understanding elliptical galaxies.

From this point onwards, various morophologies were explored.
In rapidly rotating galaxies, thin discs evolve. 
Those with lower accelerations develop central bulges.
Elliptical galaxies rotate slowly as do many dwarf galaxies.
In 1997,  Sanders \cite {Sanders} reported work on two sets of data obtained in Groningen. 
He reported on  rotation curves of 22 spiral galaxies measured in the 21 cm line of neutral hydrogen, 
together with 11 galaxies in an earlier selected sample of Begeman  et al. \cite {Begeman}. 
The Mond formula fitted the overall shape and amplitude of the 22 rotation curves. 
One free parameter was used per galaxy and a second if a bulge was present.
A commentary is given on several galaxies and a figure shows fits to 22 galaxies with three components of
the fit.  

In 1998, de Blok and McGaugh produced data testing MOND with Low Surface Brightness galaxies
\cite {deBlok}. 
After some good detective work, they found that rotation curves of 15 galaxies fitted neatly to  MOND
after making small adjustments to the inclination angles of galaxies appearing nearly side-on; this 
affected the observed luminosities.
Their Fig. 1 illustrates the Newtonian contribution from gas and stars and fits after
correction for the inclination of the disc.
This inclination needs fine tuning within the errors to conform with Milgrom's fitting function $\mu (x)$.  

In 2000, McGaugh et al. explored the Tully-Fisher relation over 5 decades of stellar masses in
galaxies \cite {McGaugh00}. 
They recognised the fact that rotational velocities depend on the number of baryons in the galaxy
after using observed HI masses for galaxies with large gas content.
This is well illustrated in the difference between their Figs. 1(a) and 1(b). 
They comment that this direct connection with the number of baryons was an argument against
a significant mass in `dark' baryons of any form.  

In 2001, Milgrom \cite {Milgrom01} grappled with the question how a Dark Matter halo could describe a thin disc.
Such discs have large orbital angular momentum. 
He followed the approach used with Beckenstein in 1984, which derives equations of motion from a Lagrangian.
Mond predicts that the `Dark halo' needs to have a disc component and a rounder component with radius-dependent
flattening, becoming spherical at large radii.
He comments that this structure is at odds with what one naturally expects from advertised halo-formation simulations.
This is at the heart of the question how different morphologies of galaxies develop in the Dark Matter scenario.

If Newton's laws apply to the Dark Matter halo, it needs to obey a Poisson equation.
This implies the structure outlined above for a thin disc. 
It will require different types of halo to describe large elliptical galaxies and dwarf spherical galaxies.
That could happen, but looks wierd.
It is a question which still continues today: how Dark Matter can reproduce the observed range of morphologies.
The problem is that the parameter $a_0$ does not appear in the Dark Matter scenario.
This is a recurrent question in papers of Sanders.

In a year 2002 paper, Milgrom \cite {Milgrom02} moves away from the asymptotic range of MOND.
He makes the point that galaxies with high central densities should show no non-Newtonian acceleration
at small radius $r$. 
This emerges later as a fundamental point.
He also makes the point that $a_0 \simeq cH_0/2\pi$, where $H_0$ is the Hubble constant and $c$ is the velocity of light.
A simple explanation of this value will be presented later.

In late 2003, Milgrom and Sanders pointed out a `Dearth of Dark Matter in Ordinary Elliptical Galaxies' \cite {MilgromSa}. 
They refer to new data of Romanovsky et al. \cite {Romanovsky} on three elliptical galaxies.
As pointed out by Milgrom in his first paper \cite {MilgromA}, the shape of MOND rotation curves depends on
${\xi} = (MG/R_e^2 a_0)^{1/2} = V^2_{\infty}/R_e a_0$, where $R_e$ is a measure of the size of the baryonic galaxy;
they take this as the half-mass radius.
Galaxies with ${\xi}\,  {\gg} 1$ have internal accelerations $\geq {a_0}$ in their main body, which is thus in the Newtonian regime.
At the other end, low surface density galaxies with ${\xi} {\ll} 1$ are in the MOND regime throughout.
This was the first time the high acceleration regime of galaxies had been probed.
Line-of-sight dispersions vary slowly, as they show in their Fig. 1. 

In 2004, McGaugh \cite {McGaugh04} carried out a careful study of disk galaxies.
In a masterly presentation of the data, he outlines what is well determined and what is not.
He concentrates on rotationally supported disk galaxies. 
The essential conclusion is that the invisible Dark Matter contribution is proportional to the visible number of baryons: 
the tail wags the dog! 
Data in the upper four panels of his Fig. 4 are close to scatter plots, whereas the bottom two show a clear correlation 
with acceleration.
On his Fig. 5, the light-to-mass ratio is tightest when the MOND prescription is used.
High surface brightness gives results close to MOND.
Only where the mass-to-light ratio falls is there large scatter.
This indicates how baryons are crucial to the interpretation of the data.  

A continuation of this point appears in a following paper of McGaugh \cite {McGaugh05}.
He comments that including gas as well as stars, the Tully-Fisher relation works even for low brightness
dwarf spherical galaxies. 
The basic point is that the Tully-Fisher relation depends on the acceleration.
He shows that Cold Dark Matter halos give a poor fit using a parametrisation close to the
Navarro, Frenk and White parametrisation \cite {Navarro}. 
In his Fig. 1, he shows that MOND gives an excellent fit to NGC 2403 of rotation curves and mass distribution.
The Dark Matter prediction is far from the data fitted by MOND.

In 2007, Milgrom and Sanders presented MOND analyses for several of the lowest mass disc galaxies
below $4 \times 10^8 M_\odot$ \cite {Milgrom07}.
They show close fits to rotation curves of 4 galaxies from the work of Begum et al. \cite {Begum}.
These galaxies are in the deep MOND regime, with low accelerations at all radii. 
They comment that the MOND result in such cases is close to a pure prediction as opposed to a one parameter fit.
Sanders and Noordermeer extend the MOND analysis to 17 high surface brightness, early-type disc 
galaxies derived from a combination of 21 cm HI lines observations and optical spectroscopy data
\cite {SandNoor}.
These are data of Noordermeer and van der Hulst in Groningen \cite {Noordermeer}.
Fits are close to data and they show the breakdown on  velocity of rotation into Newtonian, stellar and gas
components.

In 2008, Sanders and Land \cite {SandLand} made use of data of Bolton et al. \cite {Bolton} from the Sloan Lens Survey.
Whole foreground galaxies function as strong gravitational lenses to produce multiple images (the ``Einstein ring'') of 
background sources.
Bolton et al. measured the ``fundamental plane'': an observed relations between effective radius, surface brightness
and velocity dispersion, using 36 strongly lensing galaxies.
The lensing analysis was combined with spectroscopic and photometric observations of individual lens galaxies,
in order to generate a ``more fundamental plane'' based on mass surface density rather than surface brightness.
They found that this {\it mass-based} fundamental plane exhibits less scatter and is closer to expectations of the 
Newtonian virial relation than the usual luminosity based on the fundamental plane.
The conclusion is that the implied Mass/Luminosity values within the Einstein ring do not require the presence of a
substantial component of Dark Matter.
Sanders and Land show in their Figs. 1 and 2 close linear correlations between masses derived from lensing and
surface brightness, 

Milgrom himself reviewed the status of MOND at that time \cite {Milgrom08}.
He argues that MOND predictions imply that baryons alone determine accurately the full field of
each and every individual galaxy. 
He comments that this conflicts with the expectations of the Dark Matter paradigm because of the haphazard
formation and evolution of galaxies and the very different influences that baryons and Dark Matter are subject to
during their evolution, e.g. the very small baryon to Dark Matter fraction assigned by $\Lambda\, CDM$.
In MOND, all physics is predicted to remain the same under a change of units of length ${\ell \to \lambda \ell}$, of
time ${t \to \lambda t}$ and no change in mass units, $m \to m$; in words, if a certain configuration is a solution of the
equations, so is the scaled configuration.

Likewise, if $\hat {\bf r}(t) = \lambda {\bf r}(t/\lambda)$ is a trajectory where $m_i$ are at $\lambda {\bf r}_i (t/\lambda)$,
the velocities on that trajectory are ${\bf V}(t) = {\bf V}(t/\lambda)$;
i.e. a point mass remains a point mass of the same value.
Another relation is that if $m_i \to \lambda m_i$, $\bf {r}_i \to {\bf r}_i$ and $t \to \lambda ^{-1/4} t$.
So scaling all the masses leaves all trajectories unchanged, but all velocities scale as $m^{1/4}$ and accelerations
then scale as $m^{1/2}$. 
The bottom line is that rotation curves of individual galaxies are based only on observed baryonic masses.

In 2009, Stark, McGaugh and Swaters \cite {Stark} examined the Baryonic Tully-Fisher relation using gas dominated
galaxies.
They assembled a sample from 7 sources totalling $\sim 40$ low surface brightness and dwarf galaxies, which have
a high percentage of gas.
The stellar mass was not zero, so they considered a wide range of stellar population models.
Since these galaxies are gas dominated, the difference in stellar mass-to-light ratio from the different models had 
little impact. 
They were careful to select galaxies with inclinations $\geq 45^\circ$, i.e. approaching face-on. 
They checked that observed rotation curves flattened out at large radii within errors.
The conclusion is that the exponent in the data is $3.94 \pm 0.07(stat.) \pm 0.08(syst.)$, compared with the 
predicted value 4 from the Tully-Fisher relation.

In 2010, Milgrom developed a new formulation of MOND called QUMOND \cite {qumon}.
This handles, for example, a large galaxy distorting a small one.
The fundamental problem goes back to the work of Beckenstein and Milgrom in 1984 \cite {Beckenstein}.
It is not clear what the Hamiltonian or Lagrangian is that controls MOND.
Milgrom follows the idea that the MOND potential $\phi$ produced by a mass distribution $\rho$ satisfies the Poisson equation
for the modified source density. 
This is an idea which appears to work. 
He develops the algebra in Sections 3 and 4 of his paper.
It includes the constraint that the centre of mass motion of a pair of galaxies is correctly described.
This deals with the important case of perturbations between a pair of interacting galaxies.
The algebra  is actually formulated  in a general way so that it can in principle cope with interactions between many galaxies.
For a large galaxy like the Milky Way, the effects of General Relativity are at the level of $\sim 2 \times 10^{-4}$.
A following version BIMOND included the constraints of General Relativity, so as to be able to cope with effects on the
scale of the Universe \cite {Bimond}.
These two procedures provide a formalism which becomes valuable when models of galaxies develop further.

In 2011, Scarpa et al. made observations of 6 globular clusters \cite {Scarpa}.
Hernandez and Jim\' enez and Allen report a detailed study of the velocity dispersions of stars at radii of 8 globular clusters
\cite {Hernandez}. 
Like Scarpa et al., they conclude that tidal effects are significant only at radii larger by factors of 2--10 than the radius where
MOND flattens the curves.
They also show that the velocity dispersion $\sigma$ varies with the mass $M$ of the cluster as $M^{-4}$ within errors;
this is the expected analogue of the Tully-Fisher relation arising from Jeans' Law.
This result is independent of luminosity measurements used in interpreting galactic rotation curves.
In galaxies, the mass $M$ within a particular radius is not easy to determine, and is usually taken as the mass where
rotation curves flatten out. 
Further study of globular clusters is desirable. 

In 2012, McGaugh \cite {McGaugh12} reports an updated analysis of the Baryonic Tully-Fisher relation using 41 gas-rich galaxies.
My Fig. 1 shows the observed results compared with MOND.
McGaugh reports $a_0 = (1.3 \pm 0.3) \times 10^{-10}$ m s,$^{-2}$, where errors covers both
statistics and systematics.
\begin {figure}[htb]
\begin {center}
\epsfig {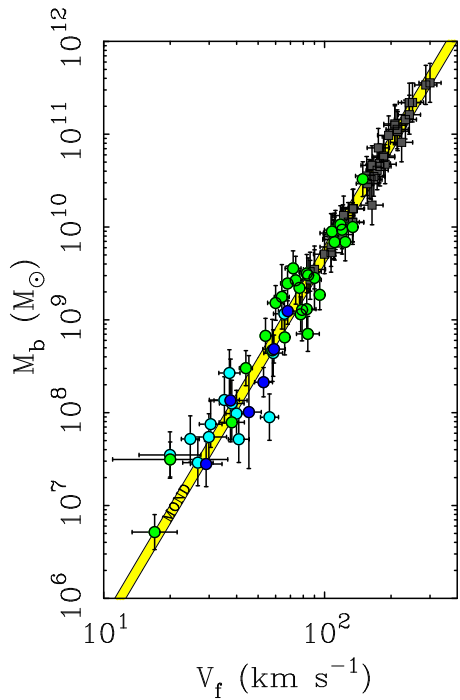}
\caption{Baryonic mass as a function of the asymptotic velocity $V_f$ of rotation curves of galaxies. The band and its width
show the expectation from MOND and detailed fits to rotation curves.} 
\end {center}
\end {figure}

Angus et al. \cite {Angus} report a useful code for calculating QUMOND. 
It solves the Poisson equation on a 2-dimensional grid.
It uses Kuzmin disks as defined by Binney and Remaine  with a surface density $\Sigma (R) = \frac {aM}{2\pi}(R^2 + a^2)^{-3/2}$.
 
Milgrom tested MOND over a wide acceleration range in X-ray ellipticals \cite {Milgrom12}. 
Two galaxies have been measured over very large galactic radii ($\sim 100$ and $\sim 200$ kpc) assuming hydrostatic
balance of the hot gas enshrouding them. 
Measured accelerations span a wide range, from ${\ge 10a_0}$ to $\sim 0.1a_0$.
He shows two figures comparing the fit to MOND with data up to unusually high masses of $12.4 M_\odot$; contributions from
stars and x-ray gas are shown.
He comments that in the context of the Dark Matter paradigm, it is unexpected that the relation between baryons and
dark matter is described  so accurately by the same formula that accounts for disc-galaxy dynamics. 

Milgrom himself summarised 10 cardinal points of MOND in December 2012 \cite {Milgromsum}

\section {New results}
In the year 2013, there have been mounting criticisms of the standard cosmological model $\Lambda \, CDM$.
It predicts that galaxies the size of the Milky Way should be accompanied by $100-600$
roughly isotropic haloes of smaller satellite galaxies formed by random fluctuations
of Dark Matter \cite {KroupaM} \cite {FamaeyM}.
This has been known since 1999. 
In an article entitled `Where are the missing galactic satellites?', Klypin et al. estimated
100-300 satellites, depending on radius, see their Fig. 4 \cite {Klypin}.
Moore et al. estimated $\sim 500$ satellites, see their Fig. 2. \cite {Moore}.
In fact, the Milky Way has $\sim 24$ satellites and Andromeda, our nearest large galaxy,
has $\sim 28$ \cite {Collins}, where one in each case could be an interloper.  A further
point is that the satellites are highly correlated in both radial and momentum phase space,
rather than being spherically distributed as Dark Matter predicts.
Milgrom \cite {qumon} is insistent that tidal effects of large galaxies have strong effects on their
satellites and is probably a key factor in determining their phase space distributions.
The best form to use for such calculations is QUMOND \cite {QUMOND}.

Yet another result comes from L\" ughausen et al. who study an unusual type of
galaxy called a `polar ring galaxy' \cite {Lughausen}.
The one they study has a small bright gas-poor disc with a central bulge, but in addition
an orthogonal gas-rich disc, referred to as a polar disc.
There are Coriolis forces between the two discs.
Observed velocities in both discs are well predicted by MOND, whereas Dark Matter
predicts a roughly spherical distribution inconsistent with the data.

McGaugh and Milgrom present two papers on Andromeda dwarfs. 
In the first paper \cite {dwarf1}, they compare recently published velocity dispersions of stars for 17
Andromeda dwarf spheroidals with estimates of MOND predictions, based on the luminosities of
these dwarfs, with reasonable stellar $M/L$ values and no Dark Matter.
The two are consistent within uncertainties.
It is necessary to take account of tidal effects due to the Milky Way on Andremeda dwarf galaxies. 
For Andromeda, only red giants can be tracked due to distance.
They predict the velocity dispersions of another 9 dwarfs for which only photometric data were available.
In the second paper \cite {dwarf2} they test their predictions against new data.
Results give reasonable stellar mass-to-light ratios, while Newtonian dynamics give large mass
discrepancies.
They comment that MOND distinguishes between regimes where the internal field of the dwarf or the
external field of the host  dominates.
The data appear to recognise this distinction, which is a unique feature of MOND,  not explicable in
$\Lambda \, CDM$.

There is a major result from Milgrom \cite {Milgrom13}.
He has studied weak gravitational lensing of galaxies using data of Brimioulle et al \cite {Brimioulle}. 
They examined foreground galaxies illuminated by a diffuse background of distant galaxies.
They remove signals from the centres of foreground galaxies so that their edges and
haloes can be studied.
Their objective was to study the Dark Halos of galaxies.
An elementary prediction of MOND is that the asymptotic form $\sqrt {a_0 g}$ of the
curvature leads to a logarithmic tail $V(r) = -\sqrt {GMa_0} \ln _e(r/r_1)$ to the Newtonian
potential; here $r_1$ is the mean radius for this term.
This tail lowers the zero point energy of the Newtonian acceleration.
Milgrom transforms this equation into the variables used by Brimioulle et al. 
I myself have checked his algebra and arithmetic and agree.  
Milgrom shows that their results obey MOND predictions accurately over a range of 
accelerations $10^{-9}\, -\, 10^{-11}\,$  m $\,$ s$^{-2}$.
Averaged over this range, results are a factor $\sim 40$ larger than predicted by conventional Dark 
Matter haloes surrounding galaxies.
 Fig. 2 shows the ratio of observed acceleration $A$ to Newtonian acceleration as a 
function of $x = -{\rm {\log }_{10}} \, g_N$.
 The curve is well known experimentally over the range Milgrom uses, but is less reliable beyond this. 
At the peak acceleration $a_0$, the effect is larger than Dark Matter predicts by a factor 
$\sim 65$.
The unavoidable conclusion is that the standard $\Lambda CDM$ model needs
serious modification.
\begin{figure}[htb]
\begin{center}
\vskip -13mm
\epsfig {file=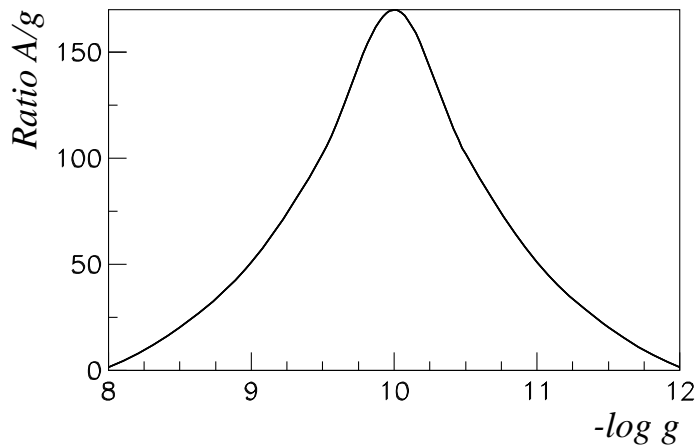,width=12cm}
\vskip -8mm
\caption {The observed ratio between total acceleration {A} and Newtonian acceleration ${g}$.}
\end {center}
\end {figure}

Fig. 3 displays Milgrom's fit to the data of Brimioulle et al. 
The MOND predictions for baryonic mass-to-light ratios 1,1.5, 3 and 6.
The measurements are reproduced from Fig. 28 of Brimioulle et al.
Triangles mark Red Galaxies and squares Blue Galaxies.
There is a difference in absolute magnitudes of velocity dispersions $\sigma$, due to different luminosities of
red and blue galaxies. 
\begin{figure}[htb]
\begin{center} 
\vskip -10mm
\epsfig{file=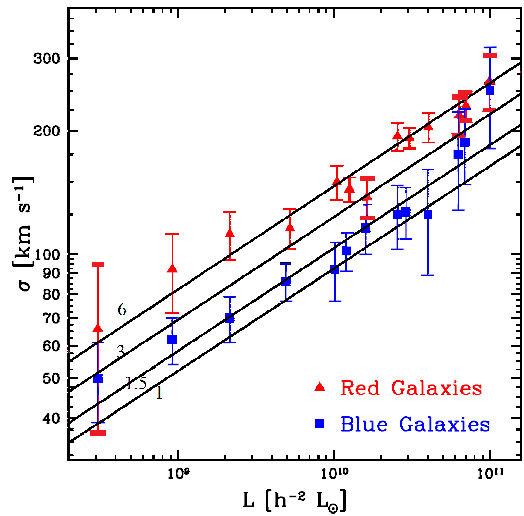,width=7cm}
\vskip -2mm
\caption{The Mond prediction for baryonic mass-to-light ratios 1, 1.5, 3 and 6. The measurements are reproduced from
Fig. 28 of Brimioulle et al.; triangles mark Red Galaxies and squares Blue galaxies.}
\end{center}
\end{figure} 

A corollary follows from Milgrom's fit to the data of Brimioulle et al. which 
agree with MOND, but are far from the prediction of $\Lambda CDM$. 
Here the photons come from distant galaxies. 
The Baryonic Acoustic Oscillations are likewise carried by photons, which in this case
originate from the Cosmic Microwave Background. 
Surely these most be treated in the same way.
The conventional assumption made in the work of Schmittful et al.  \cite {Challenor} is that photons
from the Cosmic Microwave Background are bent in weak gravitational lensing only
by Newtonian dynamics (including small corrections for General Relativity).
However, since MOND fits the data of Brimioulle {\it et al.} but $\Lambda CDM$ does not
by a large margin,  the astrophysics community should be alert to the fact that an 
additional energy $\sqrt {GMa_0} \ln  _e(r/r_0)$ originates from integrating the
acceleration $\sqrt {GMa_0}/r$;
$r_0$ is the radius where the acceleration is $a_0$.
This is needed over the range of accelerations  where MOND explains the gravitational rotation curves, Fig. 2.
It is not presently included in the fit to the Baryonic Acoustic Oscillations.
Some form factor will be needed at intergalactic distances beyond the range explored by Brimioulle et al.
For very large $r$, the red shift gradually suppresses the logarithmic term because the Universe was younger when light was emitted.
At small $r$, the mass $M$ in the formula falls in a way which needs to be fitted empirically.
A criticism levelled at MOND is that it does not fit accurately the third peak, so treating the Baryonic
Acoustic Oscillations correctly is of prime importance.

Two recent papers of Sanders make valuable reading \cite {Sanders1} \cite {Sanders2}.
In the second, he argues that `MOND, as a theory, is inherently falsifiable' and $\Lambda CDM$
is not, because of the flexibility in the way data are fitted. 
MOND has the merit that it gives a specific distribution of accelerations depending on just one parameter 
$a_0$ over the range where rotation curves of galaxies deviate significantly from Newton's 
laws.

\section {A new approach}
Up to this point, the discussion has largely concerned the peripheries of galaxies and
the approach to the asymptotic limit of the acceleration.
In Ref. \cite {cjp}, a completely different viewpoint is developed, from which the conclusion is
that quantum mechanics plays a fundamental role in forming galaxies. 
For astrophysicists this is an unfamiliar idea.  
However, from a Particle Physics viewpoint it is simple and precise.
Particle Physics governs Atomic and Particle Physics. 
It plays a fundamental role in forming Black Holes. 
Why not galaxies too?
From a Particle Physics perspective, the natural way to express gravitation is in terms
of quantised gravitons.

Before plunging into this story, it is necessary to correct one unfortunate piece of wording in the article \cite {cjp}. 
It refers to the Hubble acceleration. 
In fact the Hubble constant has dimensions of velocity.
This has no effect on the algebra or results.

The procedure used in Ref. \cite {cjp} is to adopt commonly used forms of Milgrom's $\mu$ function
to determine the non-Newtonian component of the acceleration observed at the edges of galaxies.
This peaks at or close to $a_0$ where it is bigger than $g_N$ by a large factor $\sim 171$.
This acceleration is then integrated to determine the associated energy function.
It turns out that the result fits naturally to a Fermi function with the same negative sign as
gravity.
It can be interpreted as an interaction between gravitons and nucleons (or electrons).
Fitting functions are available from the author. 
This Fermi function lowers the total energy by $0.5 GM$ at radius $r_0$ where $g_N$
reaches $a_0$; here $M$ is the mass within radius $r_0$. 
It represents an energy gap like those observed in doped semi-conductors and superconductors.

\begin{figure}[htb]
\begin{center} 
\vskip -15mm
\epsfig{file=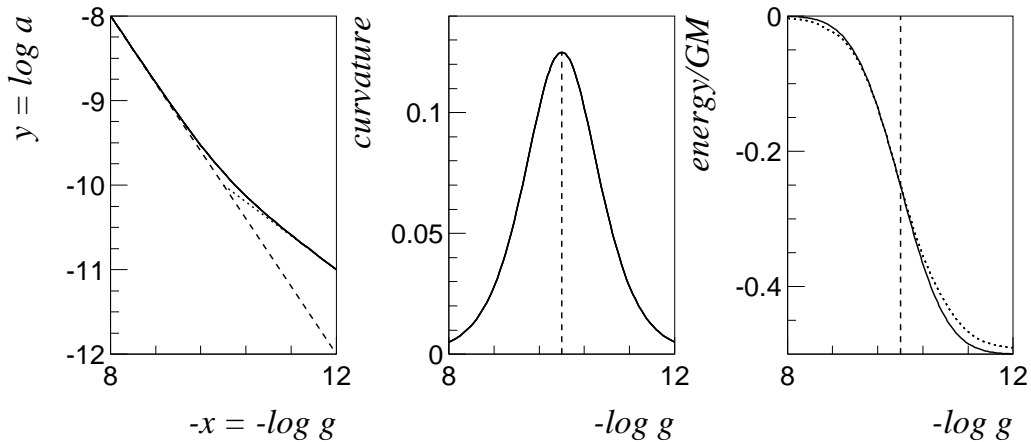,width=15cm}
\vskip -6mm
\caption
{(a) The full curve shows the result of equation (6); the dashed  
line shows $g_N$, and the dotted one a straight line given by $\sqrt {a_0g_N}$;
(b) the curvature $d^2y/dx^2$ from (a);  
(c) full curve: the energy  derived from (b); the dashed curve is discussed in
the text; vertical dashed lines mark $a_0$.}
\end{center}
\end{figure} 
There is information over the whole range of accelerations from $~10^{-7}$ m s$^{-2}$ 
upwards; at the lowest point the acceleration is almost purely Newtonian.
Results of this approach are shown in Fig. 4(a).
Data are shown on a log-log plot.
There are three functions in common use for the $\mu$ function used for MOND.
They are illustrated in Fig. 19 of the review of Famaey and McGaugh \cite {Famaey}.
The smoothest form, given by Milgrom is
\begin {equation}
\mu (x) = \sqrt {1 + 1/(4\chi ^2)} - 1/(2\chi),
\end {equation}  
where $\chi = a/a_0$.
From algebra given in eqns. (7)-(9) of Ref. \cite {cjp}, the result is
\begin {equation} 
g^2_N  + a_0g_N = a^2.
\end {equation}
This gives the full curve of Fig. 4(a).
Its curvature $d^2y/dx^2$ is a measure of the additional acceleration.
It is straightforward to derive a formula for $d^2y/dx^2$ (see \cite  {cjp});
evaluating it gives the curve shown on Fig. 4(b).
It peaks at $x = 10^{-10}$ m s$^{-2}$.
it can be approximated by a Gaussian which drops to half-height at $9.6\%$ of the value of $x$ at the peak.
The conclusion is that galaxies have considerable stability.
Note that this conclusion applies to galaxies of all sizes in view of the scaling relation used by Milgrom.

The point of interest is that on Fig. 4(a), there appears to be a cross-over between Newtonian acceleration
at low $x$ and another regime at large $x$.
Asymptotically, the total acceleration, taken from MOND, is $\sqrt{GMa_0}/r$.
Taking this as $-d\phi /dr$, where $\phi$ is a potential induced by the `extra' acceleration
\begin {equation}
\phi = -\sqrt {GMa_0}\ln (r/r_1);
\end {equation}
here $r_1$ is the mean radius of this term. 
Because $a_0 \sim 10^{-10}$, $\phi$ is very small.
However, it does explain the asymptotic straight-line at the right-hand edge of Fig. 4(a).
It also explains the long range tail observed by Brimioulle et al. 

There is an  `extra' energy in addition to Newtonian energy.
This is obtained by integrating the `extra' acceleration numerically.
From Fig. 4(a), $g_N = e^x = GM/r^2$; $e^{x/2} = \sqrt {GM}/r$, so $E_1 = -GM/r = -\sqrt {GM} e^{x/2}$.
The appropriate equation is
\begin {equation}
H_{11}\Psi _1 + V\Psi_2 = E\Psi _1;
\end {equation}
$V$ describes the `extra` energy and $E$ is the total energy.
In a quantum mechanical situation, there is a companion equation
\begin {equation}
H_{22}\Psi _2 + V\Psi _1 = E\Psi _2.
\end {equation}
These are a coupled pair of equations with two solutions
\begin {equation}
(H_{11} - E)(H_{22} - E) - V^2 = 0.
\end {equation}
This equation was first derived in 1931 by Breit and Rabi \cite {Breit}.
The same formalism describes mixing between the three neutrinos $\nu _e$, $\nu _\mu$ and $\nu _\tau$ and also 
the CKM matrix of QCD and CP violation in decays of $K^0$ mesons. 
For galaxies, classical expectation  values $<H_{11}>$ and $<H_{22}>$ are to be substituted into
Eq. (9).
The two solutions of the Breit-Rabi equation are
\begin {equation}
E = \frac {E_1 + E_2}{2} \pm \sqrt{\left( \frac {E_1 - E_2}{2} \right) ^2 + V^2}.
\end {equation}

\begin{figure}[htb]
\begin{center} 
\vskip -20mm
\epsfig{file=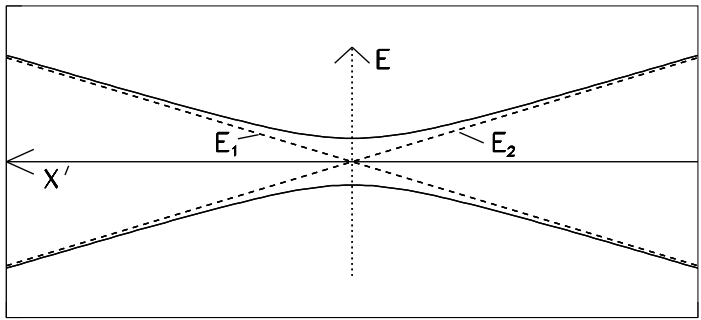,width=13cm}
\vskip -10mm
\caption {Sketch of two crossing  lines; full lines show them including mixing, dashed lines without; $E_1$ and $E_2$ 
label the convention for eigenvalues in the case of no mixing.}
\end{center}
\end{figure}
A clearer picture of what is happening is obtained by rotating Fig. 4(a) clockwise by $35.78 ^\circ$; 
this is the mean angle of the dashed line to the $x$-axis, $45^\circ$, and the angle $\tan^{-1}(0.5)$  of the
dotted line. 
The rotation of axes is the Bogoliubov-Valatin  transformation, first discovered in an obscure phenomenon in
nuclear structure physics by Bogoliubov and Valatin \cite {Bogoliubov} \cite {Valatin}.
The upper curve in Fig. 5 shows the result.
This equation arises in quantum mechanics whenever two energy levels cross as a function of $x$.
The separation between the energy levels depends on the degree of mixing governed by $V$.
 
\begin{figure}[h]
\begin{center} 
\vskip -14mm
\epsfig{file=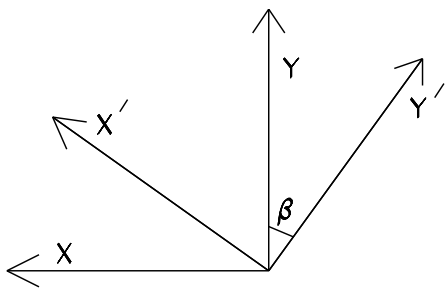,width=11cm}
\vskip -47mm
\caption {Axes $x$, $y$,\, $x'$ and $y'$.}
\end{center}
\end {figure}
Fig. 6 shows the rotation of axes; it is about the point $x = -10$, $y = -10$, where the two straight lines 
of Fig. 1(a) cross:
\begin {eqnarray}
x' &=& (x + 10) \cos \beta - (y + 10) \sin \beta \\
y' &=& (x + 10) \sin \beta + (y + 10) \cos \beta.
\end {eqnarray}
Substituting Eq. (10) gives an exact expression for the curve in $x', y'$ axes.

The full solution of the Breit-Rabi equation is given in Section 3.2 of Ref.$ \cite {cjp}$.
It is given by
\begin {eqnarray}
E_2 &=& \sqrt {GM}\epsilon (x) \\
V &=& \sqrt {GM}W(x),
\end {eqnarray}
where $W$ is given by the standard form of the Fermi function:
\begin {equation}  
W(x) \propto \left[ 1 + \exp \left(
\frac {E - E_F}{\beta E_F}\right) \right]^{-1} ;
\end {equation}
$E_F$ is the energy at the centre of the Fermi function and $\beta$ is a
fitted constant.
The value of $E_2$ describes the asymptotic variation of energy and is given by $\sqrt {a_0g_N}$.
The depth of the Fermi function is $-0.5\, GM$.
The magnitude of this term can be traced to the factor 2 difference
in slope of $g_N$ and that of the asymptotic form $\sqrt {a_0 g_N}$.
A Bose-Einstein condensate does not fit the data, hence demonstrating that
the condensate is not in the gravitational field itself.
 
Let us now return to Fig. 4(c).
Here there is a slight complication. 
Fig. 4(b) is the acceleration measured in $x,y$ axes. 
However, in $x',y'$ axes the curvature increases by a factor 1.48.
In addition, there is a small visible displacement of the centre of curvature
in Fig. 4(a) by an offset of $-0.203 \times 10^{-10}$ in $x$. 
What then emerges from the Breit-Rabi equation is that 
$g_N$ is rather small near $x = a_0$ compared with that originating from 
the extra acceleration $dW/dx'$.
This second term dominates by a large factor $\sim 171$ at the 
centre of the curve.
This ratio falls by $50\%$ at $x=-10.6$, to 30 at $x=-11$, then -6.0 at 
$x=-11.5$ and  $\sim 1$ at $x=-12$.  
Results for the `extra' acceleration are symmetric about $a_0$ except for 
the term $\sqrt {a_0g_N}$ of Eq. (10).
The conclusion is that the curved part of Fig. 4(a) is the dominant 
feature near $x = a_0$ where Newtonian gravitation is a rather small 
perturbation.

Consider the effect of this result near the centre of the Fermi function at
$E = E_F$ in Fig. 4(c). 
If we retain only the dominant terms in $W$ and $dW/dx$, 
\begin {eqnarray} 
dE/dx &\to (GM/{\ln _e 10}) 2W/dx \\
E &\to ( {GM/\ln _e 10}) 2W.
\end {eqnarray}
Apart from the factor $\sqrt {GM}/\ln _e 10$, $dE/dx$ may be interpreted as
the modulus of a Breit-Wigner resonance with $x$-dependent width:
\begin {equation}
 BW = \frac {\Gamma (x)/2}{E - E_F - i\Gamma (x)/2}.
\end {equation}
The energy starts at zero because there is no difference between Newtonian
energy and total energy at the top of Fig. 3(c), and its central energy is
shifted downwards by $0.25 GM$.
In Ref. \cite {cjp}, the effect of using alternative forms of Milgrom's $\mu$ function
were tested.
Although acceleration curves change significantly, as shown by the dotted curve of Fig. 4(c), the Fermi function 
is affected only at the ends of the range $x = 8$ to $12$ by at most $\pm 4\%$.

The Fermi function acts as a funnel, attracting gas and dust into the periphery of
the galaxy.
The shape of the Breit-Wigner can be alternatively expressed as an Airy function
with a modest form factor. 
It comes from the coherent interactions of gravitons with nucleons over a large
volume.
The form factor can arise for example from supernovae which heat considerable volumes.
Such effects resemble defects like those observed in superconductors.

What about the missing lower branch of the Breit-Rabi equation?
On this branch both $W$ and $dW/dx'$ change sign.
The change of sign requires that this branch describes an excited
state rather than a condensate. 
(Remember that energies of both gravity and $W(x)$ are negative). 
Such an excited state is likely to decay on a time scale much less than
that of galaxies, so this branch has not been observed.

It is interesting that phenomena appear on a log-log scale.
The logarithmic dependence has a natural explanation in terms of the statistical mechanics
of the interaction between gravitons and nucleons.
Schr\" odinger shows in a delightfully simple approach that quantum mechanics requires
the logarithmic dependence for both Fermi-Dirac statistics and Bose-Einstein  \cite {Schrodinger}. 
This is further direct evidence for quantum mechanics at work, since it is a purely quantum
phenomenon.
A fit using Bose-Einstein statistics fails completely to fit galactic data.

A prediction is that in Voids there will be no Fermi function lowering the energy.
It is observed that many large galaxies appear at the edge of the Local Void \cite {Nusser}. 
This can occur by attracting gas and dust out of the neighbouring Void. 
The converse occurs in clusters of galaxies. 
Each galaxy in a cluster has a Fermi function and this results in complex interferences
between individual galaxies in the cluster and general lowering of the energy. 
This may account for the fact that MOND falls short by a factor of 2 in predictions of 
accelerations in galaxy clusters. 
 
In Ref. \cite {cjp} a further conjecture is made about a connection to Dark Energy. 
Experiment tells us that in galaxies, the asymptotic form of the
acceleration is $\sqrt {a_0g_N}$.
This leads to the question: what governs the asymptotic acceleration? 

If MOND successfully models the formation of galaxies and globular
clusters, it raises the question of how to interpret Dark Energy.
In a de Sitter universe, the Friedmann-Lema\^ itre-Robertson-Walker model (FLRW)  
smoothes out structures using a $\Lambda CDM$ function which models the 
gross features.
My suggestion is that galaxies create fine-structure and the FLRW model should include
into Dark Energy the sum over these structures. 
This sum increases as galaxies grow in the recent past.
It has the potential to account for the late-time acceleration of the Universe.
This remains to be tested.

The way in which the Bogoliubov-Valatin transformation works in nuclei has been reviewed recently in a paper
of Ring \cite {Ring}. 
It is intricate, but is well described by Ring.
It depends on the spontaneous violation of symmetries such as rotational symmetry in deformed nuclei and
the gauge theory in superfluid systems.
The phenomenon is called `backbending'.
In nuclei, more than one basis state exists and there are towers of resonances separated by two units of spin,
made from each of these basis states.
At large excitations they cross in a similar way to the two regimes in Fig. 7(a) here.
Basically what happens is that the excited states can decay via emission of photons and this damps
the excited states as a function of energy, see Fig. 2 of \cite {Ring}. 
There is then quantum mechanical mixing amongst the excited states.

\end {document}